\documentclass[reprint,superscriptaddress,amsmath,amssymb,aps,prl]{revtex4-2}
\usepackage{bm}
\usepackage{float}
\usepackage{graphicx}
\usepackage{mathtools}
\usepackage{braket}
\usepackage[usenames,dvipsnames]{color}
\usepackage[normalem]{ulem}
\usepackage[svgnames]{xcolor}
\usepackage{bm}
\usepackage{multirow}
\usepackage[utf8]{inputenc}
\usepackage{array}
\usepackage[colorlinks,linkcolor=blue,citecolor=blue,urlcolor=blue]{hyperref}
\usepackage{tabularx}
\usepackage{booktabs}
\usepackage{makecell}
\usepackage{comment}

\begin{document}

\title{Observation of thermal Hall effect in diamond}

\author{Marcin Matusiak}
\email[e-mail: ]{m.matusiak@intibs.pl}
\affiliation{Institute of Low Temperature and Structure Research, Polish Academy of Sciences\\ ul. Okólna 2, 50-422 Wrocław, Poland}

\author{Andrzej Ptok}
\affiliation{Institute of Nuclear Physics, Polish Academy of Sciences, \\ ul. W. E. Radzikowskiego 152, 31-342 Krak\'{o}w, Poland}

\author{Maria Szlawska} 
\affiliation{Institute of Low Temperature and Structure Research, Polish Academy of Sciences\\ ul. Okólna 2, 50-422 Wrocław, Poland}

\author{Kamran Behnia}
\email[e-mail: ]{kamran.behnia@espci.fr}
\affiliation{Laboratoire de Physique et d'\'Etude de Mat\'{e}riaux (CNRS)\\ ESPCI Paris, PSL Research University, 75005 Paris, France}

\date{\today}

\begin{abstract}
 Numerous insulators, including non-magnetic ones, have been unexpectedly found to display a finite thermal Hall signal, which has stimulated debate about its origin. Here, we report on a study of the thermal Hall effect in two diamond single crystals. The transverse thermal conductivity ($\kappa_{xy}$) was found to peak at a temperature close to that at which the longitudinal thermal conductivity ($\kappa_{xx}$) reaches its maximum. The measured $\kappa_{xy}$, with an amplitude of 340~Wm$^{-1}$K$^{-1}$ at $B=10$~T, is the largest ever observed, while the $\kappa_{xy}/ \kappa_{xx}$ ratio follows the phenomenological trend identified in other insulators. Our observation implies the existence of an intrinsic thermal Hall effect in a generic phonon gas. We argue that the rough amplitude of both the thermal Hall angle and the thermal Hall resistivity can be accounted for by simple arguments invoking fundamental constants and quantum-mechanical constraints on solid state cohesion.

\end{abstract}
\maketitle

\section{Introduction}

The origin of the electric Hall effect, known since the end of the nineteenth century, is firmly established. An electric charge ($q$) moving in a magnetic field (\textbf{\textit{B}}) is subject to the magnetic component of the Lorentz force, ${\bm F} = q ( {\bm v} \times {\bm B})$, where ${\bm v}$ denotes the velocity. This causes its trajectory to curve, provided that ${\bm v}$ and ${\bm B}$ are not parallel. As ${\bm F} = 0$ for $q = 0$, the motion of charge-neutral particles is not expected to be affected by a magnetic field, and hence phonon transport is usually assumed to be $B$-independent. It was therefore believed that the thermal Hall effect, which involves the appearance of a transverse thermal gradient ($\nabla_y T$) in response to a longitudinal heat current ($Q_x$) under an applied field ($B_z$), occurs exclusively in electrically conductive materials~\cite{Zhang2000,Matusiak2005}.

However, this paradigm had to be revised following the observation of the thermal Hall effect in the insulating terbium gallium garnet~\cite{Strohm2005}. This compound, Tb$_3$Ga$_5$O$_{12}$, exhibits pronounced magneto-elastic coupling, which was invoked in early accounts of its thermal Hall response~\cite{Sheng2006, Araki2008}. It was later found that the thermal Hall effect occurs in many other insulators belonging to a variety of distinct families~\cite{Ideue2017,Grissonnanche2020,Boulanger2020,Akazawa2020,Chen2022,
Uehara2022,Sim2021,Jiang2022,Chen2024,Chen2024-2,Ataei2024,sharma2024phonon,
Li2025,Meng2024,lishi2025,Sugii2017,Li2020,Li2023}, including non-magnetic ones~\cite{Sugii2017,Li2020,Li2023}. This experimental observation stimulated numerous theoretical studies
\cite{Qin2012,Chen2020,Flebus2022,Guo2022,Mangeolle2022} invoking either an extrinsic or an intrinsic origin for the phonon thermal Hall effect (PTHE).

The recent observation of PTHE in elemental insulators (black phosphorous~\cite{Li2023}, silicon~\cite{lishi2025} and germanium~\cite{lishi2025}) indicates that PTHE may be a common (or 'universal'\cite{lishi2025}) phenomenon in any phonon gas, like the Senftleben--Beenakker (SF) effect in molecular gases. The latter, discovered and understood in the previous century~\cite{beenakker1970}, refers to the influence of a magnetic field on the transport property of any gas consisting of non-spherical molecules, including the emergence of an off-diagonal thermal response~\cite{hermans1970transverse}.

The analogy between a phonon gas and a real gas of molecules or atoms is a recurrent feature in textbooks of condensed matter physics~\cite{ashcroft1976solid,hook2013solid}. The SF effect implies that neutral particles can couple to a magnetic field. Nevertheless, it was largely forgotten in the discussions of PTHE. It is instructive to contrast the origin of the Hall effect in electronic and in molecular systems. For electrons, the Hall signal arises because a magnetic field bends the trajectory of electrons between collisions. For molecules, it arises due to the influence of the magnetic field on collisions. Neutral phonons are more likely to resemble neutral molecules than charged electrons.

The absence of diamond among insulators hitherto investigated constitutes a significant gap in our knowledge. In Si~\cite{Middelmann2015}, Ge~\cite{McCammon1963} and black P~\cite{SUN20162098}, the thermal expansion coefficient anomalously changes sign. In contrast, diamond has a small and positive thermal expansion coefficient~\cite{Stoupin2010} and is arguably the closest solid to the harmonic approximation. The observation of the thermal Hall effect in diamond would conclusively rule out any exotic origin. 

Given its exceptionally large thermal conductivity, however, detecting a finite $\kappa_{xy}$ is an experimental \textit{tour de force}, requiring a high resolution for measuring sub-millikelvin transverse temperature differences.

\begin{figure*}[ht!]
\centering
\includegraphics[width=0.8\linewidth]{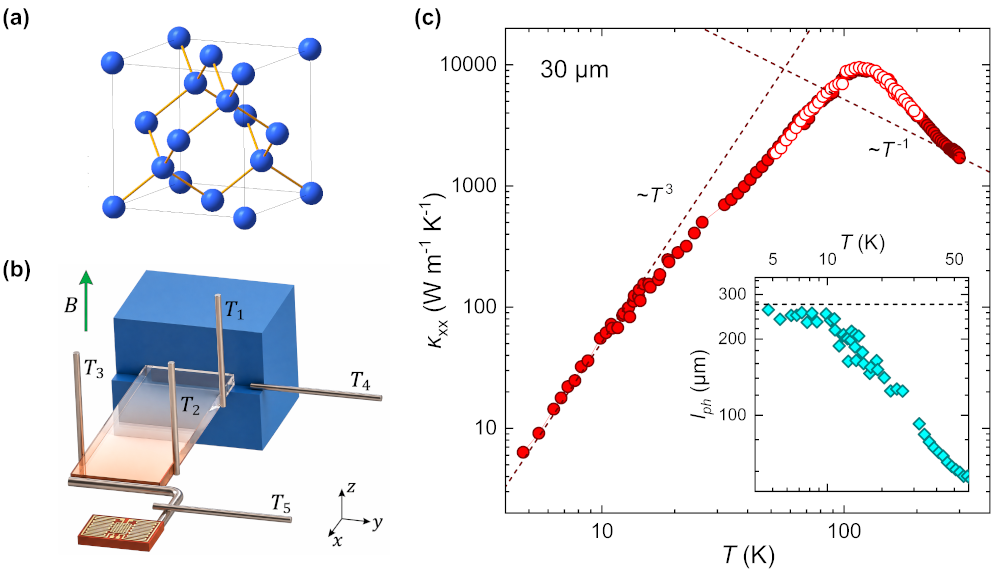} 
\caption{\textbf{Longitudinal thermal transport in diamond.} (\textbf{a}) The face-centered cubic crystal structure of diamond. (\textbf{b}) Schematic setup for measuring longitudinal and transverse thermal conductivities. (\textbf{c}) Temperature dependence of the longitudinal thermal conductivity $\kappa_{xx}$ at zero field (filled circles) and in a magnetic field of $14$~T (open circles). The inset presents the temperature dependence of the calculated phonon mean free path ($l_{ph}$). At low temperature $l_{ph}$ approaches the limit of $274$~$\mu\text{m}$ (dashed line) estimated on the basis of the sample size.}
\label{fig.1}
\end{figure*}

Here, we report on a study of thermal transport in diamond single crystals subject to a magnetic field. By finding a thermal Hall signal with features shared by other insulating crystals, we demonstrate that this is a generic phenomenon driven by the influence of a magnetic field on the interaction between phonons. We then argue that the rough amplitude of the peak thermal Hall angle and the thermal Hall resistivity can be accounted for. 

\section{Results}

Figure~\ref{fig.1}(a) shows the schematic face-centered cubic crystal structure ($Fd\bar{3}m$) of diamond, and Laue diagrams (which are shown in Fig.~S1 in the Supplementary Information~\cite{SM}) confirmed that the samples were oriented with the $[1\ 1\ 0]$ direction along the $x$ axis, $[-1\ 1\ 0]$ along the $y$ axis, and $[0\ 0\ 1]$ along the $z$ axis.

In the experimental setup, shown schematically in Fig.~\ref{fig.1}(b), the heat current was induced along the $x$ axis, the transverse temperature gradient was measured along the $y$ axis, and a magnetic field was applied along the $z$ axis. As expected, we found a very large longitudinal thermal conductivity ($\kappa_{xx}$), reaching a maximum at $T=115$~K with a value of $\kappa_{xx} = 9200$~W\,m$^{-1}$\,K$^{-1}$, in the $30$~$\mu$m-thick sample. This is in reasonable agreement with what was reported for the purest natural single crystals of diamond (type IIa)~\cite{Pan2013,Berman1976}. The application of a $14$~T magnetic field does not affect the values of $\kappa_{xx}$ (shown in Fig.~\ref{fig.1}(c) as open points) within our experimental margin. The dashed lines in Fig.~\ref{fig.1}(c) indicate a low-temperature $T^3$ dependence~\cite{Graebner1992}, expected for boundary scattering, whereas at high temperature the data approach the $1/T$ dependence characteristic of dominant Umklapp processes~\cite{Peierls1955}. The inset in Fig.~\ref{fig.1}(c) presents the phonon mean free path ($l_{ph}$) calculated using the relation $\kappa_l = \frac{1}{3} C_v v_s l_{ph}$. Assuming reasonably that $C_v \approx C_p$ at low temperature, we used $C_p(T)$ data extrapolated with a $T^3$ function~\cite{Desnoyers1958}. For $v_s$ we used $12.5$~km/s, which was reported for the transverse acoustic phonon branch~\cite{Ward2009}. This population is expected to carry almost 90\% of heat at low temperatures~\cite{Inyushkin2023}. The phonon mean free path in the Casimir limit~\cite{Casimir1938} was calculated from the thickness ($t$) and width ($w$) of the sample: $l_{ph}= 2 \sqrt{w~t / \pi}$ . 

\begin{figure}[!t]
\centering
\includegraphics[width=1.0\linewidth]{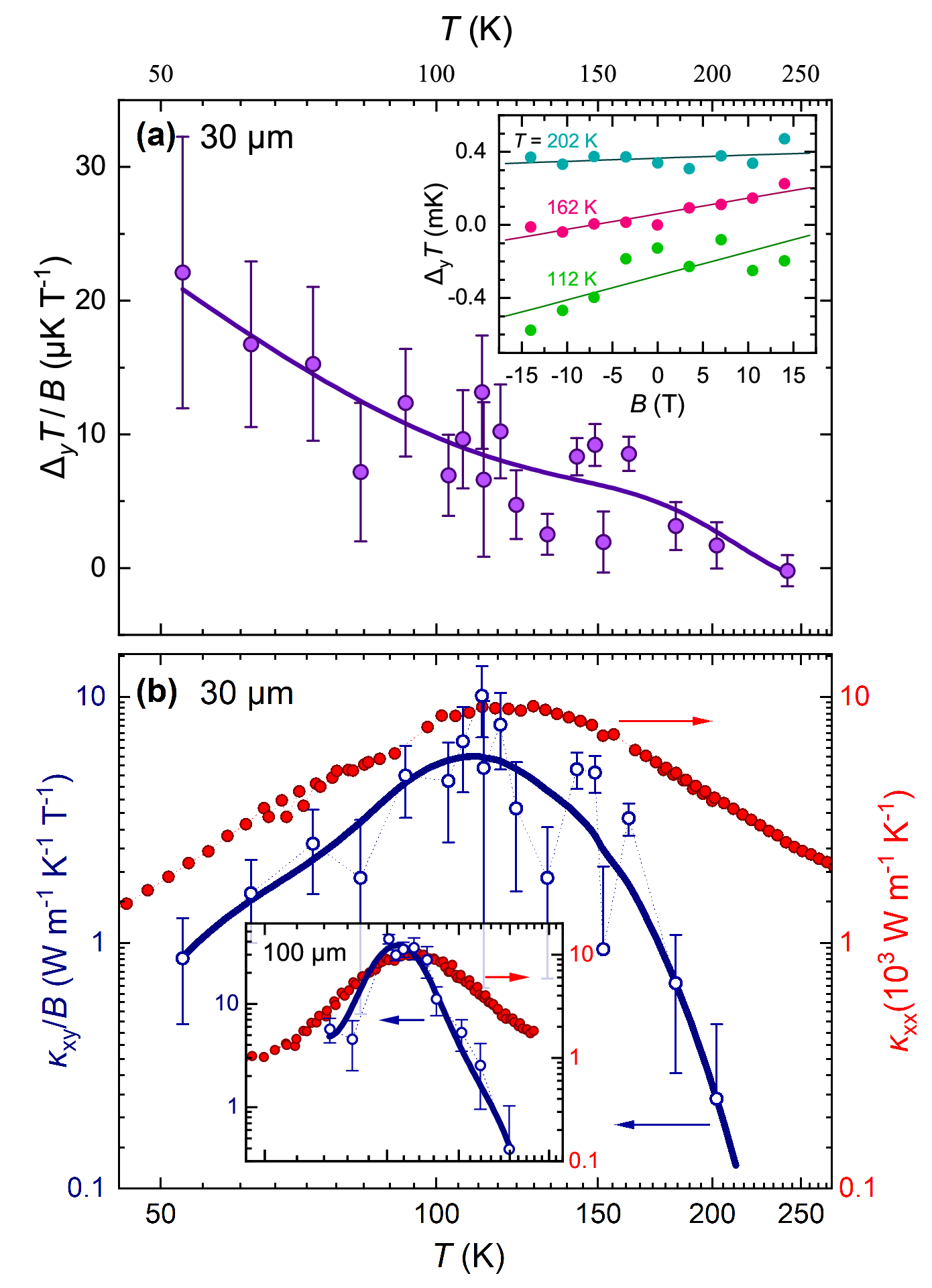} 
\caption{\textbf{Transverse thermal conductivity in diamond.} (\textbf{a}) Temperature dependence of the slope of $\Delta_y T(B)$ in 30~$\mu\text{m}$ thick diamond. The solid dark-violet line is a guide to the eye. The inset shows the field variation of $\Delta_y T$ for three example temperatures. The $\Delta_y T$ data at each temperature have been vertically shifted for clarity. (\textbf{b}) The temperature dependence of the transverse thermal conductivity plotted along the corresponding $\kappa_{xx}(T)$ for the 30~$\mu\text{m}$ sample. Inset presents the respective results for the 100~$\mu\text{m}$ sample in the same temperature range. The solid dark-blue lines are guides to the eye.}
\label{fig-02}
\end{figure}

We succeeded in resolving a finite $\Delta_y T$ induced by magnetic field, as presented in Fig.~\ref{fig-02}(a). The relatively large error bars result from the extremely high thermal conductivity of diamond that effectively suppresses the measured signal according to the relation $\nabla_y T = \frac{\kappa_{xy}}{\kappa_{xx}}\, \nabla_x T$. Furthermore, the longitudinal thermal gradient is also small for the same reason. As a result, at temperatures around the maximum of $\kappa_{xx}$, a heater power of $3.2$~mW generated a signal of several mK for the longitudinal temperature difference and a few tenths of a mK for the transverse temperature difference (at $B = 15~\mathrm{T}$). The raw data values of $\Delta_y T$, measured at different $B$ values, are presented in the inset in Fig.~\ref{fig-02}(a) for three distinct temperatures and show that the field-odd component dominates the $\Delta_y T(B)$ signal, which is close to linear. The very small slope of $\Delta_y T(B)$, which at $T \approx 50$~K reaches a value of approximately $20$~$\mu$K/T and decreases with increasing temperature, illustrates the scale of the experimental challenge.

\begin{figure}[!t]
\centering
\includegraphics[width=1\linewidth]{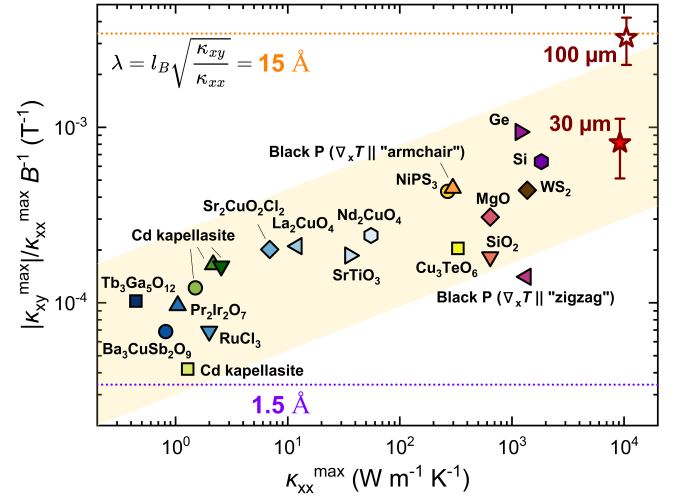} 
\caption{\textbf{The peak thermal Hall angle in diamond and in other insulators.} The ratio between peak values of $|\kappa_{xy}|/B$ and $\kappa_{xx}$ plotted versus $\kappa_{xx}^{\mathrm{max}}$ for various insulators~\cite{Li2020,Grissonnanche2020,Boulanger2020,Akazawa2020,Chen2022,Uehara2022,Lefran2022,Li2023,Meng2024,lishi2025,Boulanger2022,Sugii2017,Hirokane2019,
ling2026phononthermalhalleffect,Guo2026}. The dotted lines at the bottom and top denote $|\kappa_{xy}|/\kappa_{xx}~B^{-1}$ calculated for $\lambda = 1.5~\mathrm{\AA}$ and $15~\mathrm{\AA}$, respectively.}
\label{fig.03}
\end{figure}

The effect observed in the $30$~$\mu$m-thick sample was reproduced by measurements on another sample with a thickness of $100$~$\mu$m, for which the corresponding $\kappa_{xy}(T)/B$ dependence is shown in the inset of Fig.~\ref{fig-02}(b). 
In both samples, $\kappa_{xy}/B$ exhibits a pronounced maximum, reaching $7.5$~W\,m$^{-1}$\,K$^{-1}$\,T$^{-1}$ for the $30$~$\mu$m sample and $34$~W\,m$^{-1}$\,K$^{-1}$\,T$^{-1}$ for the $100$~$\mu$m sample. 
Although the absolute magnitude differs between the two samples, the presence of a maximum in both cases demonstrates that the observed transverse response is robust. These values are unrivalled among previously reported materials.

The ratio of transverse-to-longitudinal thermal conductivities normalised by magnetic field, $(|\kappa_{xy}|/\kappa_{xx}) B^{-1}$ peaks at $0.8$ ($2.7$) $\times 10^{-3}$~T$^{-1}$ in the $30$ ($100$)~$\mu\text{m}$ samples. This can be assimilated to the thermal Hall angle, quantifying the field-linear misalignment between the heat flux and the thermal gradient. Figure~\ref{fig.03} compares the amplitude of this quantity in diamond with what was reported in other crystalline insulators. Plotting the maximum $(|\kappa_{xy}|/\kappa_{xx}) \, B^{-1}$ as a function of maximum $\kappa_{xx}$ shows that they are not proportional to each other, but there is a visible positive correlation. \mbox{A~$10^4$-fold} increase in the magnitude of maximum $\kappa_{xx}$ is accompanied by a tenfold increase of the thermal Hall angle. In other words, across different crystals, there is a rather weak positive correlation between the thermal Hall angle and the maximum phonon mean free path. Moreover, for all materials,  the length scale scattered from the Hall angle, calculated as $\lambda = \ell_B\sqrt{\frac{\kappa_{xy}}{\kappa_{xx}}}$, falls within  1.5 -- $15~\mathrm{\AA}$. Here $\ell_B=\sqrt{\frac{\hbar}{eB}}$ is the magnetic length. 

\begin{figure}[!b]
\centering
\includegraphics[width=1\linewidth]{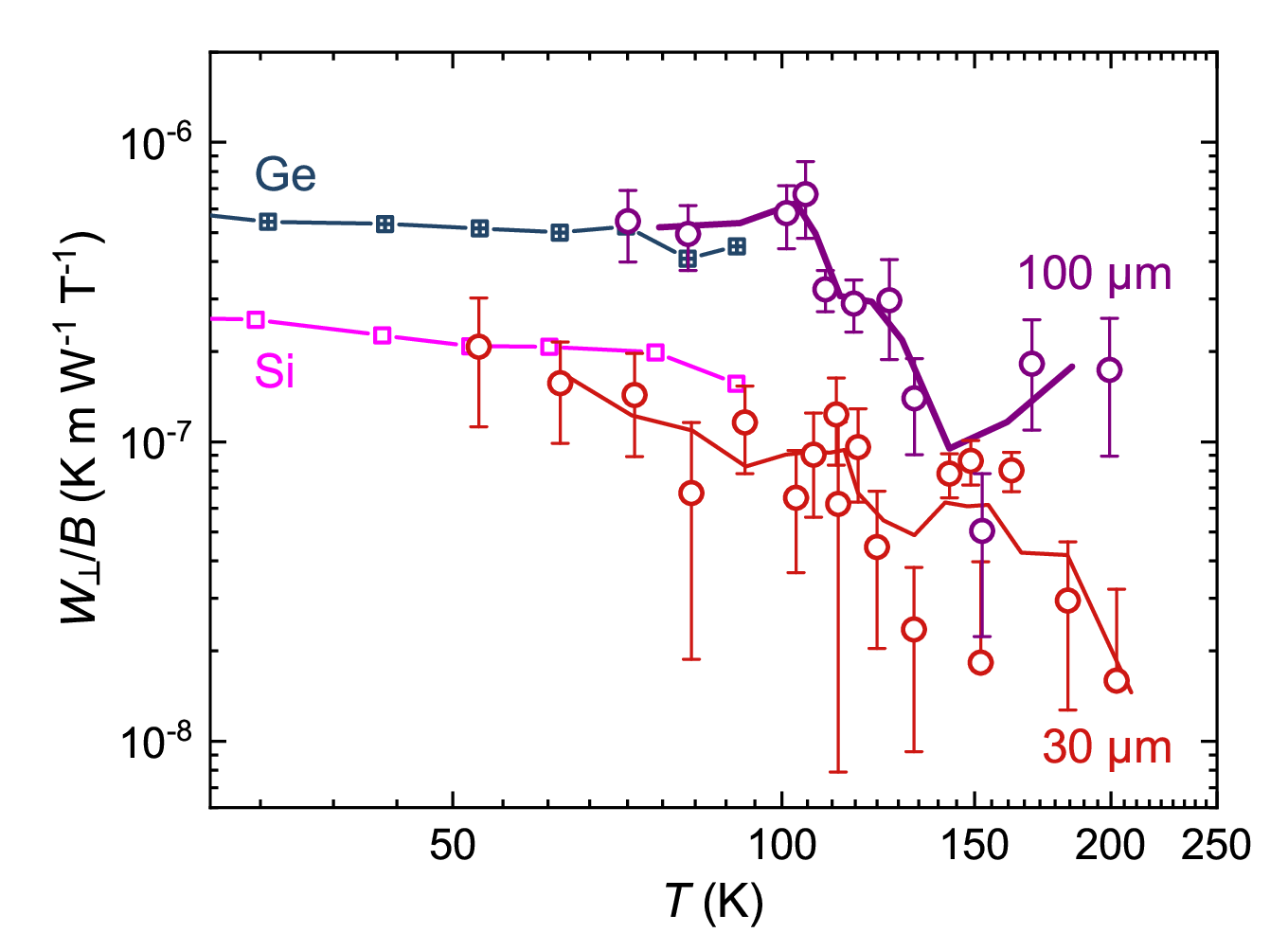} 
\caption{\textbf{Transverse thermal resistivity in diamond compared with Ge and Si} The temperature dependence of the transverse thermal resistivity $W_{\perp}=\nabla_yT/q_x$ divided by magnetic field for the 30~$\mu\text{m}$ and 100~$\mu\text{m}$ thick samples of diamond compared with Si and Ge~\cite{lishi2025}.}
\label{fig.04}
\end{figure}

Finally, Fig.~\ref{fig.04} shows the temperature dependence of the inverse effect, namely the thermal Hall resistivity, $W_{\perp}=\nabla_yT/q_x$. According to our data, $W_{\perp}$ tends to decrease with increasing temperature, in contrast to what was found in Si~\cite{lishi2025}, Ge~\cite{lishi2025}, black P~\cite{Guo2026} and quartz (SiO$_2$)~\cite{ling2026phononthermalhalleffect}. Nevertheless, as seen in the figure, $W_{\perp}$ in diamond has an order of magnitude comparable to what was observed in Si and Ge. In all three cases, \mbox{$W_{\perp}=\nabla_yT/q_x \, B^{-1}$} is in the range of \mbox{10$^{-7}$--10$^{-8}$ K\,m\,W$^{-1}$\,T$^{-1}$}.

\section{Discussion}

Our observation of a finite thermal Hall conductivity in diamond confirms that this phenomenon is generic to crystalline insulators. In line with previous studies of such materials, we find that the $\kappa_{xy}$ and $\kappa_{xx}$ peak at nearly the same temperature. At their common maximum, $\kappa_{xy}/\kappa_{xx}~B^{-1}$ ratio in diamond is $\simeq 10^{-3}$ T$^{-1}$. 

These findings impact the ongoing quest to identify the origin of PTHE.

First of all, the observation of a $\kappa_{xy}$ in diamond with an amplitude easily dwarfing what has been seen before implies the existence of an intrinsic source of thermal Hall effect, surviving in the absence of any exotic carriers or excitations including chiral phonons.

Second, any intrinsic scenario is expected to explain several generic features. One is that the thermal Hall signal maximises at the boundary between a high-temperature regime (where phonons scatter mostly with each other) and a low-temperature regime (where phonons are mostly scattered by the boundaries). This indicates that phonon-phonon interactions play a crucial role, confirming the suspected link with the origin of the thermal Hall response in molecular gases.

In the absence of a rigorous microscopic theory, let us try to answer two questions regarding the order of magnitude of the measured signals. {\it i)} Why the thermal Hall angle peaks at $\simeq 10^{-3}$ T$^{-1}$; {\it ii)} Why the order of magnitude of the thermal Hall resistivity normalised by magnetic field is $W_{\perp} B^{-1} \simeq 10^{-7}$~K\,m\,W$^{-1}$\,T$^{-1}$? 

Given that the order of magnitude of these measured numbers does not vary much across solids, it is tempting to look for constraints imposed by fundamental constants~\cite{Trachenko02102021}. Interatomic distances in solids provide an excellent example. They are always in the range of a few angstroms, because the Bohr radius, defined by fundamental constants, imposes a lower boundary. Another example is the melting temperature of solids, which never exceeds a Rydberg. Yet another example is the sound velocity, a few kilometers per second in any solid, a bound imposed by fundamental constants~\cite{Trachenko2020}. These examples motivate us to look for an account of the order of magnitude of the field-induced misalignment between heat flux and temperature gradient and that of the thermal Hall resistivity.

\begin{figure}[!t]
\centering
\includegraphics[width=1
\linewidth]{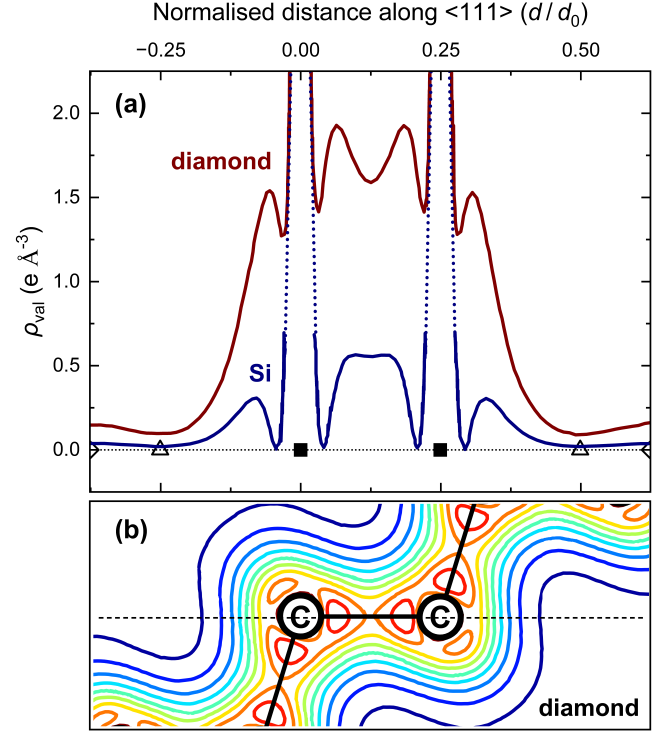} 
\caption{\textbf{Valence charge density in silicon and in diamond} ({\bf a}) Ab initio calculated static valence charge density ($\rho_{\mathrm{val}}$) of diamond and silicon in the $\langle 111 \rangle$ direction, where the diagonal length $d_0=\sqrt{3}a$ ($a$ is the lattice constant), is used to normalise the distance. Solid squares denote atomic positions, empty triangles denote the empty tetrahedral interstitial sites, and empty half-diamond symbols at the left and right margins denote the (empty) hexagonal interstitial sites~\cite{Lu1993}. ({\bf b}) The corresponding contour plot with steps of $0.20$~e/\AA.}
\label{fig.5}
\end{figure}

Phonons are neutral. Yet, they exist thanks to bonds between atoms, which can be assimilated to strings with Coulombic force constants. The non-uniformity of charge distribution in diamond-type crystals is remarkably complex: valence electrons pile up between the atoms to form localised covalent bonds. The successful Bond Charge Model~\cite{Martin1969,Weber1977} locates positively charged atomic cores at the normal lattice sites linked by Bond Charges (BCs) along the lines connecting them. As shown in Fig.~\ref{fig.5}, the electronic charge distribution in diamond-structure crystals, established with an impressive accuracy in the last decade of the previous century~\cite{Lu1993}, is remarkably complex. Looking at them, it is hard to see how lattice vibrations would avoid jumbling electric charge and coupling with magnetic field.

One route towards an intrinsic phonon Hall effect is non-adiabaticity associated with a geometric phase beyond the Born--Oppenheimer approximation \cite{Saito2019,Saparov2022,chen2025magicnonlocalgeometricforce,Behnia2025}. According to a recent proposal~\cite{Behnia2025}, in the presence of magnetic field in an anharmonic crystal, a mismatch between the Aharanov--Bohm phase of positive and negative charges emerges by a collectice atomic vibration. In this scenario, the breakdown of the Born--Oppenheimer approximation in a finite magnetic field~\cite{Schmelcher1988,Yin1994} corresponds to an area in real space, defined by the product of two length scales, the phonon wavelength and the crest atomic displacement and to be compared by the square of the magnetic length.

In diamond, the thermal Hall angle peaks close to $100$~K. The wavelength of a transverse phonon at this temperature can be estimated using $\lambda_{\mathrm{ph}} (T)= \frac{h v_t}{k_B T}$, which yields $\simeq 6$~nm. The maximum atomic displacement by atomic vibrations $\delta u_m$, at this temperature can be estimated using the Lindemann criterion, stating that a solid melts when the atomic displacement becomes as large as a significant fraction of the interatomic distance, $a$ (i.e. $\delta u_m (T_m)=c_L\sim 0.3 a$). Diamond melts at $3820$~K and its average interatomic distance is $a=0.18$~nm. Given that $\delta u_m \propto T^{-0.5}$~\cite{Khrapak2020}, this yields $\delta u_m \simeq 0.03$nm. The product of these two length scales, $\lambda_{\mathrm{ph}}$ and $\delta u_m$, quantifies the area in real space affected by a collective atomic vibration~\cite{Behnia2025}. In the present case, $2\frac{\lambda_{\mathrm{ph}}\delta u_m}{\ell_B^2} \simeq 5.4\times 10^{-4}$~T$^{-1}$, smaller than the observed signal, but within its order of magnitude. 

Let us now consider the thermal Hall resistivity. What causes $W_{\perp}/B$ to be $\simeq 10^{-7}$~K\,m\,W$^{-1}$\,T$^{-1}$? Why not a thousand times larger or smaller? A heat flux of $q_x$ applied along an axis of a crystal will generate a flow of phonons. A fraction of this phonon flow suffers resistive collisions and produces entropy on its trajectory. Energy, in contrast to entropy, is conserved. Therefore, the elastic energy density of the crystal, $\rho_U$ will have a drift velocity, $v_d$ sustained by the flow of phonons. Conservation of energy implies that $v_d=q_x/\rho_U$. The elastic energy is sustained by charged ions and bond. Let us designate its the associated charge density by $\rho_Q$. Then ,in the presence of a magnetic field, there will be a Lorentz force density of $f_L= \rho_Q v_dB$. 

Textbooks assert that ``magnetic forces do no work''~\cite{griffiths2017introduction} or ``the magnetic field does no work, since the magnetic force is perpendicular to the velocity''~\cite{jackson_classical_1999}. Indeed, the work done by the Lorentz force is $\delta W_l=F \, \delta l= (B \times v) \, v \, dt=0$. Therefore, the flux of thermal \textit{energy} cannot be deflected by such a force. On the other hand, the flow of phonons can be deflected and this would produce a transverse temperature gradient and a transverse entropy flux. Thus, what counters the Lorentz force is an entropic force with a density of $f_{th}=\rho_S \nabla_yT$, where $\rho_S$ is the entropy flux. Cancellation of the two forces leads to:
\begin{equation}
\rho_S\nabla_yT =\frac{q_x}{\rho_U}\rho_Q B
\label{thermal-Hall-res-1}
\end{equation}
This yields an expression for transverse thermal resistivity: 
\begin{equation}
W_{\perp}/B = \frac{\rho_Q}{\rho_U \rho_S}. 
\label{thermal-Hall-res-2}
\end{equation}
This formula for $W_{\perp}/B$ is similar, but more general, than what was put forward recently~\cite{ling2026phononthermalhalleffect,Guo2026}. It links $W_{\perp}/B$ to the ratio of the three densities. Assuming that $\frac{\rho_Q}{\rho_S} \approx \frac{e}{k_B}$ and that $\rho_U \approx 100~\mathrm{GPa}$, a typical value for the elastic constants and bulk modulus of solids (itself imposed by the combination of the Rydberg and Bohr radius), one obtains $W_{\perp}/B \approx 10^{-7}$~K\,m\,W$^{-1}$\,T$^{-1}$, comparable in the order of magnitude to the observed signal.

Diamond has a significantly larger bulk modulus and smaller lattice constant than silicon~\cite{Yin1981}. As a consequence, its $\rho_U$ is roughly 4.3 times larger than that of silicon. On the other hand, because of its smaller unit cell, its $\rho_Q$ is also 3.5 times larger than that of silicon. According to Eq.~\ref{thermal-Hall-res-2} their $W_{\perp}/B$ should be of the same order of magnitude. Within the present experimental margin, this appears to be the case.

Thus, even in the absence of a microscopic theory, the order of magnitude of the observed signal can be accounted for using general arguments based on conservation laws and quantitative constraints imposed by fundamental constants. 

In summary, we measured an exceptionally high transverse thermal conductivity in single-crystalline diamond and found that $\kappa_{xy}(T)$ and $\kappa_{xx}(T)$ exhibit maxima at the same temperature. By quantifying the thermal Hall angle and the transverse thermal resistivity, and comparing them with other insulators we found common features in the order of magnitude but differences in detail. Finally, we showed that the order of magnitude of both can be accounted for by simple arguments invoking fundamental constants.

\section{Methods}

Two plate-like samples of single-crystalline diamond were purchased from Diamond Materials. The first sample was $3.5$~mm long, $2$~mm wide, and $30$~$\mu$m thick, whereas the corresponding dimensions of the second were $3.5$~mm, $2$~mm, and $100$~$\mu$m. For the sake of clarity, some results regarding the second sample are included in the Supplementary Information~\cite{SM}.

The orientation of the diamond crystals was checked by the Laue backscattering technique using a Proto Laue-COS system.

The measurements were carried out in a Quantum Design PPMS system equipped with a $14$~T superconducting magnet. The sample holder was made of sterling silver, and the thermal contacts were made of $100$~$\mu\text{m}$ pure silver wires attached to the sample using DuPont 4929N silver paste. The thermal gradient was induced using a $5$~k$\Omega$ Micro-Measurements strain-gauge.

Due to the high thermal conductivity of diamond, the measurements were carried out in two steps. First, we measured the longitudinal thermal conductivity in the experimental setup schematically presented in Fig.~\ref{fig.1}(b). This involved measuring the base temperature using a Cernox thermometer and two temperature differences using type E (Constantan--Chromel) differential thermocouples calibrated in a magnetic field~\cite{Matusiak2026}. The two mentioned temperature differences were \mbox{$\Delta_x T = T_2 - T_1$} along the sample, and \mbox{$\Delta_{bh}T = T_5 - T_4$} (between the base and heater).

The latter allowed us to precisely determine the actual temperature of the sample. Second, we measured $\Delta_y T = T_2 - T_3$ (across the sample) and $\Delta_{bh}T = T_5 - T_4$ to determine the transverse temperature difference arising under an applied magnetic field: $\kappa_{xy} = \frac{\nabla_y T}{\nabla_x T}\, \kappa_{xx}$, where $\nabla_x T$ and $\nabla_y T$ are the longitudinal and transverse thermal gradients, respectively. Further details regarding the data processing are provided in the Supplementary Information~\cite{SM}.

\section{Data availability}
The data that support the findings of this study are available from the corresponding authors upon request.

\bibliography{main}

\section{Acknowledgments}

This work was supported financially by the National Science Centre (Poland) under Research Grant No.~2025/59/B/ST3/00232.

\section{Author information}

\noindent {\bf Contributions:}\\
Conceiving and performing the experiment: M.M.; conceptualisation: K.B and M.M.; Laue characterisation: M.S.; preparing the manuscript: M.M., K.B., A.P., M.S. All authors reviewed and approved the final version of the manuscript.

\noindent {\bf Corresponding authors:}\\
Correspondence to Marcin Matusiak and Kamran Behnia

\section{Ethics declarations}

\noindent {\bf Competing interests:}\\
The authors declare no competing interests.


\clearpage
\newpage

\onecolumngrid

\begin{center}
  \textbf{\Large Supplementary Information}\\[.3cm]
  \textbf{\large Observation of thermal Hall effect in diamond}\\[.3cm]
  Marcin Matusiak$^{1}$, Andrzej~Ptok$^{2}$, Maria Szlawska$^{1}$, Kamran Behnia$^{3}$ \\[.2cm]
  {\itshape
${}^{1}$Institute of Low Temperature and Structure Research, Polish Academy of Sciences, \\[.1cm] ul. Okólna 2, 50-422 Wrocław, Poland \\[.1cm]
${}^{2}$Institute of Nuclear Physics, Polish Academy of Sciences, \\[.1cm] ul. W. E. Radzikowskiego 152, 31-342 Krak\'{o}w, Poland \\[.1cm]
${}^{3}$Laboratoire de Physique et d'\'Etude de Mat\'{e}riaux (CNRS), \\[.1cm] ESPCI Paris, PSL Research University, 75005 Paris, France \\[.1cm]
  }
  (Dated: \today)
\\[0.3cm]
\end{center}

\setcounter{equation}{0}
\renewcommand{\theequation}{S\arabic{equation}}
\setcounter{figure}{0}
\renewcommand{\thefigure}{S\arabic{figure}}
\setcounter{section}{0}
\renewcommand{\thesection}{S\arabic{section}}
\setcounter{table}{0}
\renewcommand{\thetable}{S\arabic{table}}
\setcounter{page}{1}

\begin{figure*}[ht!]
\centering
\includegraphics[width=0.5\linewidth]{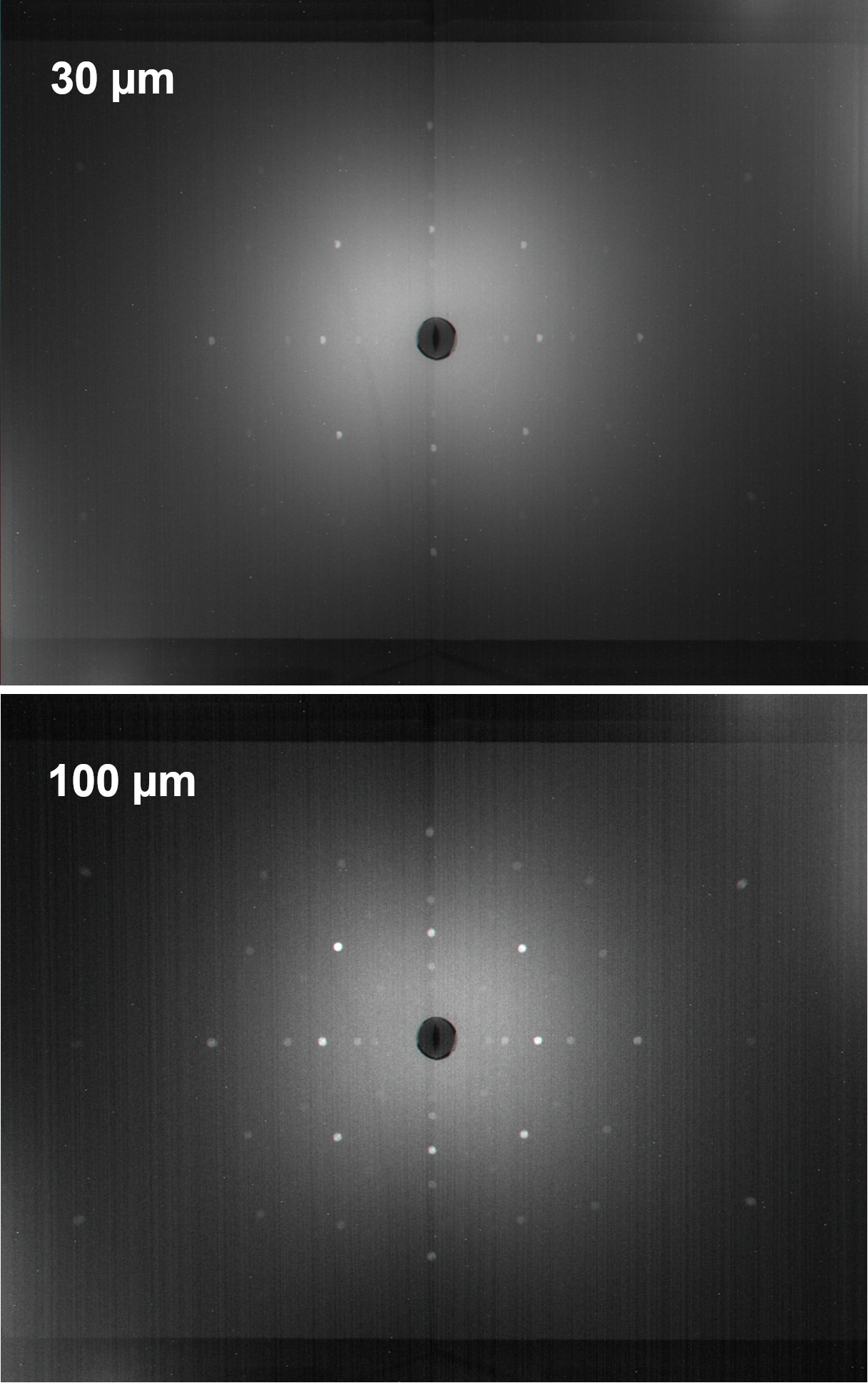} 
\caption{\textbf{Sample orientation.} Laue diffraction patterns recorded for the 30 and 100 $\mu\text{m}$ thick diamond single crystals (top and bottom panels, respectively). The X-ray beam was oriented perpendicular to the crystal surface ($xy$ plane) and parallel to the crystallographic [001] direction. The patterns confirm the $[1\ 1\ 0]$ and $[-1\ 1\ 0]$ orientations along the $x$ and $y$ axes, respectively. The reflections from the 30 $\mu\text{m}$ sample are weaker because of the smaller thickness of the plate and, consequently, the reduced diffracting volume.}
\label{fig.s1}
\end{figure*}
\begin{figure*}[ht!]
\centering
\includegraphics[width=0.9\linewidth]{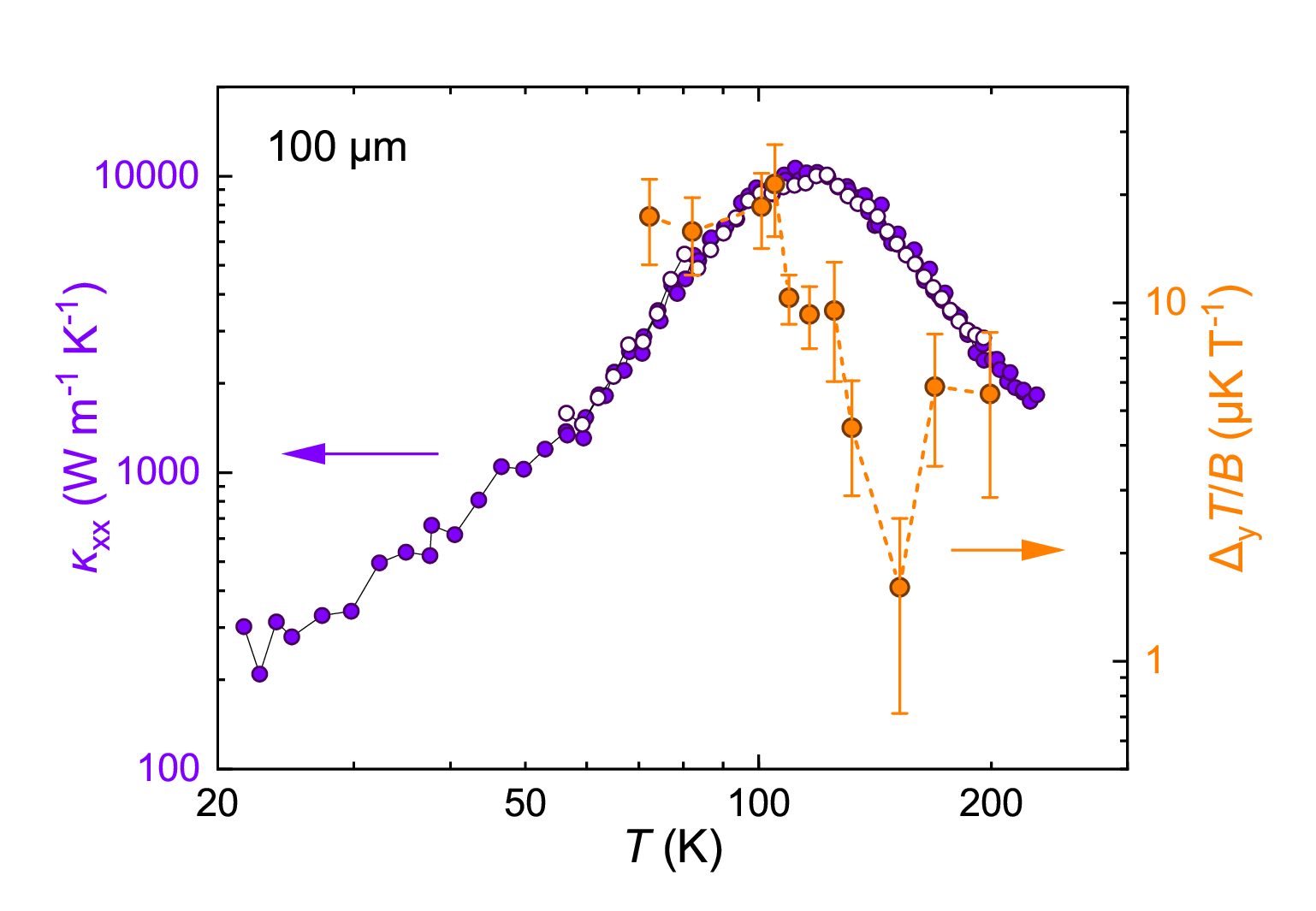} 
\caption{\textbf{Thermal conductivity of the 100~$\mu\text{m}$ thick diamond.} Left axis: the temperature dependence of the longitudinal thermal conductivity $\kappa_{xx}$ for the 100~$\mu\text{m}$ thick diamond at zero field (filled circles) and in a magnetic field of 14 T (open circles). Right axis: the temperature dependence of the slope of $\Delta_y T(B)$ for the 100~$\mu\text{m}$ thick diamond.}
\label{fig.s2}
\end{figure*}
\begin{figure*}[ht!]
\centering
\includegraphics[width=0.9\linewidth]{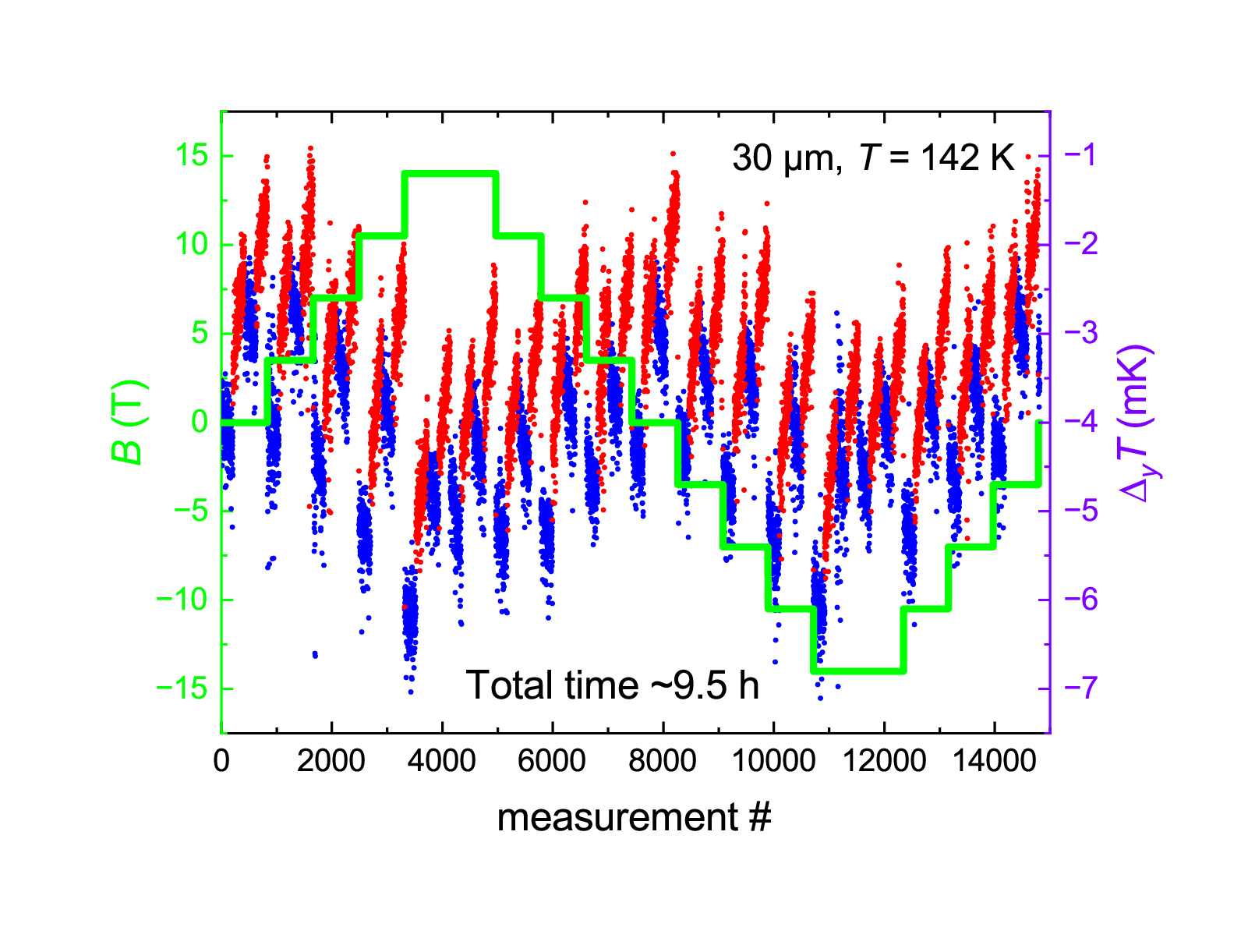} 
\caption{\textbf{The data collected for the 30~$\mu\text{m}$ thick diamond sample during a 9.5 hour long run at $T \approx 142 \mathrm{~K}$.} Left axis: the magnetic field value (green solid line). Right axis: the transverse temperature difference $\Delta_y T(B)$ with the heater ON ($\Delta_y T^{\mathrm{ON}}$, red points) and OFF ($\Delta_y T^{\mathrm{OFF}}$, blue points).}
\label{fig.s3}
\end{figure*}
\begin{figure*}[ht!]
\centering
\includegraphics[width=0.9\linewidth]{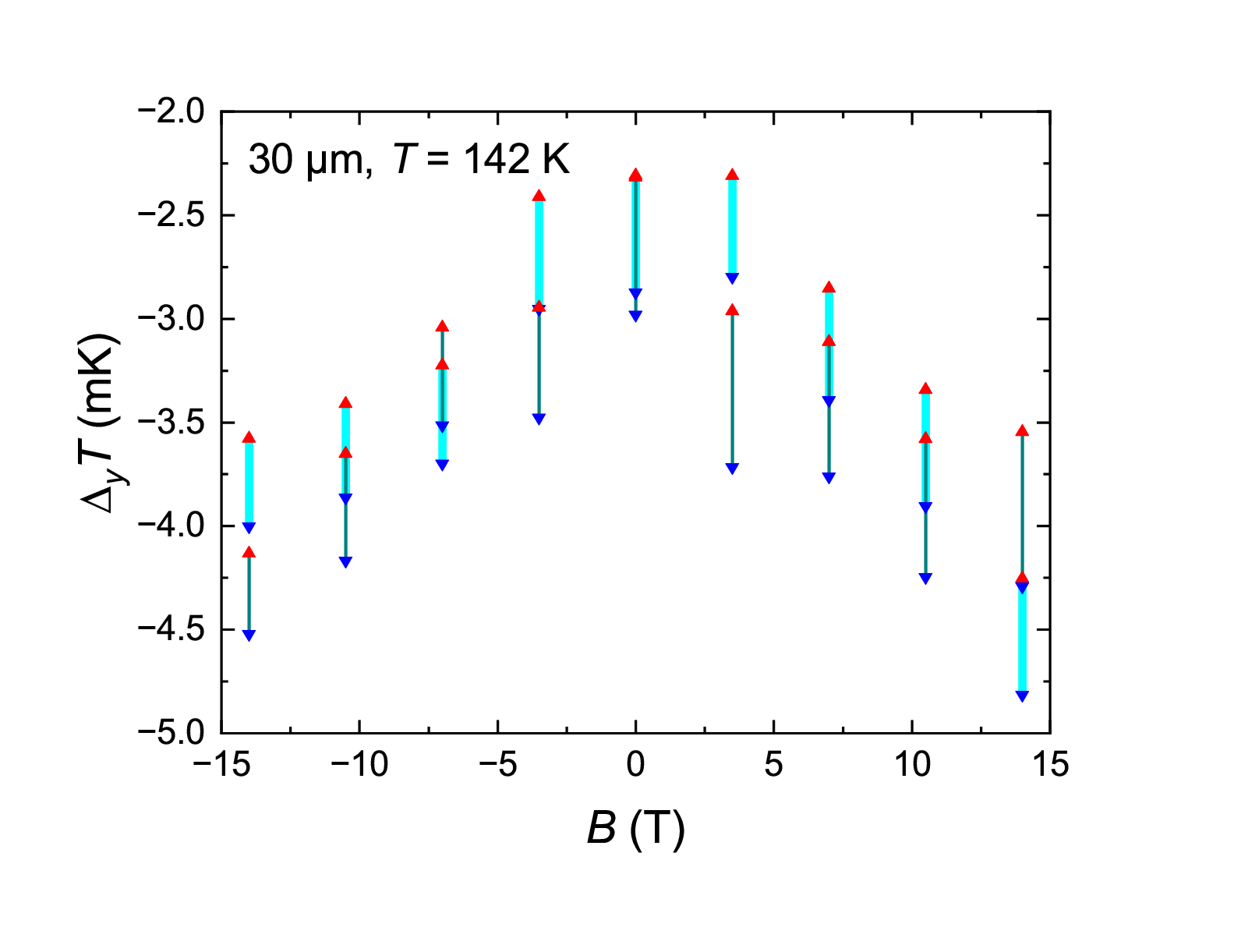} 
\caption{\textbf{The average value of the transverse temperature difference $\Delta_y T(B)$ data calculated separately for the heater in the ON (red triangles) and OFF (blue triangles) state.} Dark and light cyan lines denote a distance between $\Delta_y T^{\mathrm{ON}}$ and $\Delta_y T^{\mathrm{OFF}}$ for each field.}
\label{fig.s4}
\end{figure*}
\begin{figure*}[ht!]
\centering
\includegraphics[width=0.9\linewidth]{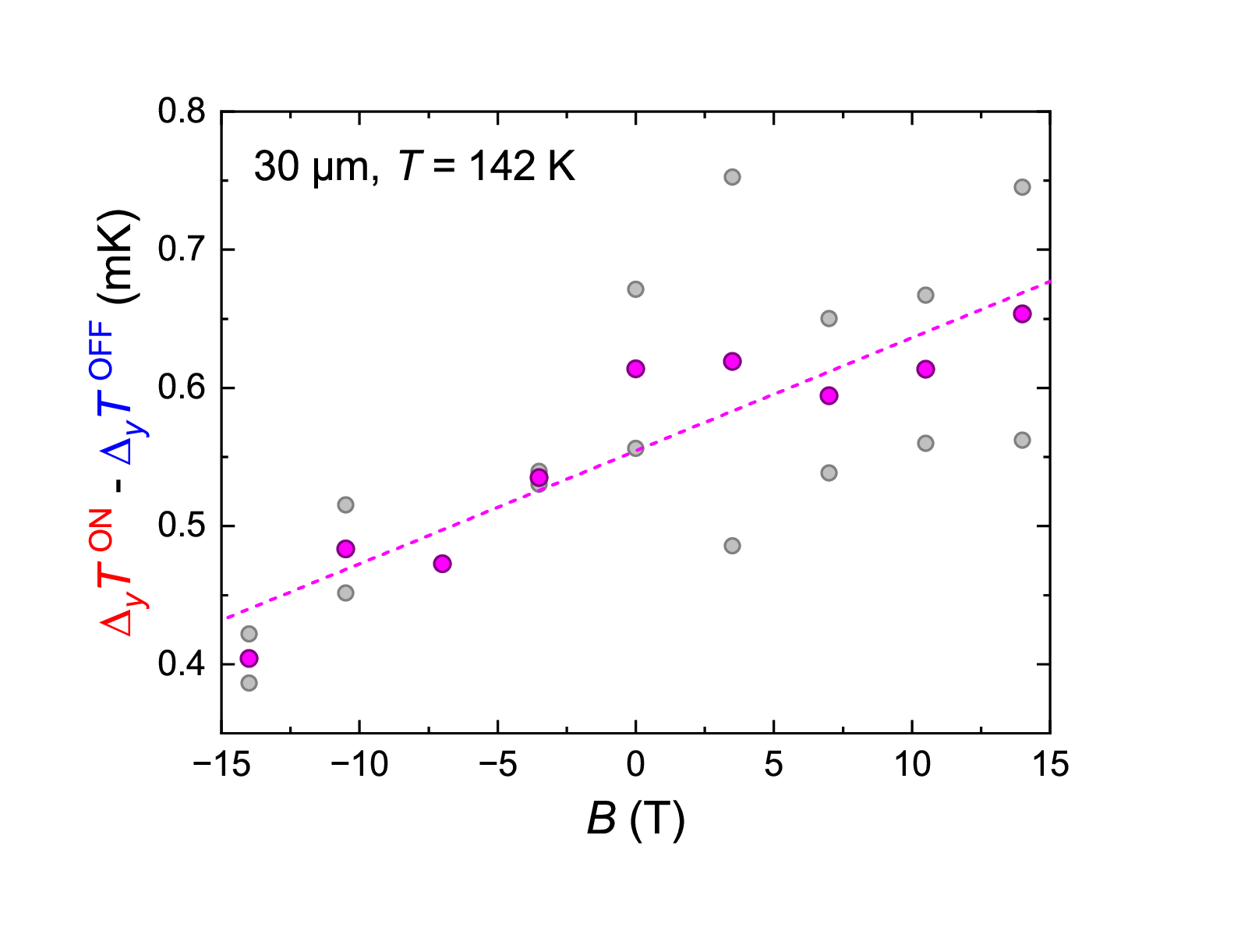} 
\caption{\textbf{The magnetic field dependence of the difference $\Delta_y T^{\mathrm{ON}} - \Delta_y T^{\mathrm{OFF}}$.} Calculated values are shown as grey points, whereas magenta points denote the field-binned averages. The dashed magenta line represents the linear fit.}
\label{fig.s5}
\end{figure*}

In this Supplementary Information, we present additional results:
\begin{itemize}
  \item Fig.~\ref{fig.s1} -- Laue diagrams
  \item Fig.~\ref{fig.s2} -- Longitudinal thermal conductivity for $100$~$\mu\text{m}$
  \item Figs.~\ref{fig.s3},~\ref{fig.s4}, and~\ref{fig.s5} -- Data processing
\end{itemize}

\newpage
\section*{Note 1. Sample orientation}


\newpage
\section*{Note 2. Thermal conductivity of the 100~$\mu\text{m}$ thick diamond.}

\newpage
\section*{Note 3. Data processing.}

\end{document}